\documentclass[english]{paper}
\usepackage[T1]{fontenc}
\usepackage[latin9]{inputenc}
\usepackage{amsmath}
\usepackage{stmaryrd}
\usepackage[numbers]{natbib}
\usepackage[all]{xy}

\makeatletter
\newcommand{\lyxaddress}[1]{
	\par {\raggedright #1
	\vspace{1.4em}
	\noindent\par}
}
\ifx\proof\undefined
\newenvironment{proof}[1][\protect\proofname]{\par
	\normalfont\topsep6\p@\@plus6\p@\relax
	\trivlist
	\itemindent\parindent
	\item[\hskip\labelsep\scshape #1]\ignorespaces
}{%
	\endtrivlist\@endpefalse
}
\providecommand{\proofname}{Proof}
\fi

\@ifundefined{date}{}{\date{}}
\usepackage{geometry}
\usepackage[all]{xy}
\usepackage{tikz}
\usetikzlibrary{arrows}

\makeatother

\usepackage{babel}
\begin{document}
\title{Context-independent mapping and free choice are equivalent: A general
proof}
\author{Ehtibar N.\ Dzhafarov}
\maketitle

\lyxaddress{Purdue University, USA, ehtibar@purdue.edu}
\begin{abstract}
Free choice (or statistical independence) assumption in a hidden variable
model (HVM) means that the settings chosen by experimenters do not
depend on the values of the hidden variable. The assumption of context-independent
(CI) mapping in an HVM means that the results of a measurement do
not depend on settings for other measurements. If the measurements
are spacelike separated, this assumption is known as local causality.
Both free choice and CI mapping assumptions are considered necessary
for derivation of the Bell-type criteria of contextuality/nonlocality.
It is known, however, for a variety of special cases, that the two
assumptions are not logically independent. We show here, in complete
generality, for any system of random variables with or without disturbance/signaling,
that an HVM that postulates CI mapping is equivalent to an HVM that
postulates free choice. If one denies the possibility that a given
empirical scenario can be described by an HVM in which measurements
depend on other measurements' settings, free choice violations should
be denied too, and vice versa. 

\end{abstract}

\section{Introduction}

The historical context for our analysis is set by a discussion that
A. Shimony, M.$\,$A. Horne, and J.$\,$F. Clauser had with J.$\,$S.
Bell \citep{BellvsSHC1985}. The subject of this discussion was the
assumptions underlying derivation of Bell's famous inequality \citep{Bell1966}
(or its offshoots \citep{CHSH1969}) for Bohm's version of the Einstein-Podolsky-Rosen
(EPR) experiment \citep{Bohm1957}. Bell originally assumed that this
derivation follows from the principle of local causality alone. According
to this principle, if outcomes of an experiment are modeled by a function
that contains as its arguments a ``hidden'' random variable and
some parameters, these parameters cannot contain spacelike remote
experimental settings. Shimony, Horne, and Clauser pointed out to
Bell that, in addition to local causality, one had to postulate that
the hidden random variable in the hypothetical model does not in any
way correlate with experimental settings, local or remote. Bell agreed
with this criticism (see \citep{Norsen2011} for a more detailed historical
description). One consequence of the present paper is that Bell did
not need to agree. He could have explained instead that one cannot
accept local causality without free choice, because denying free choice
is equivalent to denying local causality. 

Let us define the terminology systematically. For our purposes, all
empirical scenarios are described by \emph{systems of random variables}
representing measurements,
\begin{equation}
\mathcal{R}=\left\{ R_{q}^{c}:c\in C,q\in Q,q\Yleft c\right\} .\label{eq:system}
\end{equation}
Here, $R_{q}^{c}$ represents the outcome of measuring property $q$
in \emph{context} $c$ (belonging to corresponding sets $Q$ and $C$),
and $q\Yleft c$ means that $q$ is measured in $c$. A context $c$
is any set of systematically recorded circumstances under which the
properties are measured. As an example, the following is a system
of random variables describing an EPR/Bohm experiment:

\begin{equation}
\begin{array}{|c|c|c|c||c|}
\hline R_{1}^{1} & R_{2}^{1} &  &  & c=1\\
\hline  & R_{2}^{2} & R_{3}^{2} &  & c=2\\
\hline  &  & R_{3}^{3} & R_{4}^{3} & c=3\\
\hline R_{1}^{4} &  &  & R_{4}^{4} & c=4\\
\hline\hline q=1 & q=2 & q=3 & q=4 & \textnormal{system }\mathcal{R}_{4}
\\\hline \end{array}\,.\label{eq:1}
\end{equation}
Here, $q=1$ and $3$ are settings used by Alice, and $q=2$ and 4
are settings used by Bob. Contexts in (\ref{eq:1}) are defined by
the four combinations of Alice's choice and Bob's choice. In the canonical
version of the experiment, Alice's measurements are spacelike separated
from Bob's choices, and vice versa. 

A hidden variable model (HVM) of system $\mathcal{R}$ is a representation
of the variables $R_{q}^{c}$ in any given context $c$ in the form
\begin{equation}
\left\{ R_{q}^{c}\right\} _{q\Yleft c}\overset{d}{=}\left\{ \alpha\left(q,c,\Lambda^{c}\right)\right\} _{q\Yleft c},\label{eq:most general}
\end{equation}
where $\left\{ expression\right\} _{q\Yleft c}$ is a compact way
of writing $\left\{ expression:q\in Q,q\Yleft c\right\} $, with $c$
fixed. The symbol $\overset{d}{=}$ means ``is distributed as,''
$\alpha$ is some function, and $\Lambda^{c}$ are ``hidden random
variables.'' The equation says that the joint distribution of all
$R_{q}^{c}$ in a given context $c$ is the same as the joint distribution
of the corresponding $\alpha\left(q,c,\Lambda^{c}\right)$. In system
(\ref{eq:1}), e.g., we have
\begin{equation}
\begin{array}{c}
\left\{ R_{1}^{1},R_{2}^{1}\right\} \overset{d}{=}\left\{ \alpha\left(q=1,c=1,\Lambda^{1}\right),\alpha\left(q=2,c=1,\Lambda^{1}\right)\right\} ,\\
\left\{ R_{2}^{2},R_{3}^{2}\right\} \overset{d}{=}\left\{ \alpha\left(q=2,c=2,\Lambda^{2}\right),\alpha\left(q=3,c=2,\Lambda^{2}\right)\right\} ,\\
\left\{ R_{3}^{3},R_{4}^{3}\right\} \overset{d}{=}\left\{ \alpha\left(q=3,c=3,\Lambda^{3}\right),\alpha\left(q=4,c=3,\Lambda^{3}\right)\right\} ,\\
\left\{ R_{4}^{4},R_{1}^{4}\right\} \overset{d}{=}\left\{ \alpha\left(q=4,c=4,\Lambda^{4}\right),\alpha\left(q=1,c=4,\Lambda^{4}\right)\right\} .
\end{array}
\end{equation}
We can also present this example in the form of a graph, where $q$
and $q'$ are the two settings used in a given context:

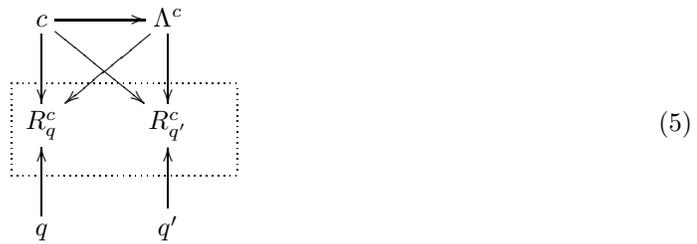
\begin{equation}
\vcenter{\xymatrix@C=1cm{c\ar[d]\ar[dr]\ar[r] & \Lambda^{c}\ar[dl]\ar[d]\\
\save"2,1"."2,2"*+++[F.]\frm{}\restore R_{q}^{c} & R_{q'}^{c}\\
q\ar[u] & q'\ar[u]
}
}
\end{equation}
An arrow here indicates that the distribution of its terminal node
may change with the value of its initial node, and the dotted box
indicates that the random variables within it are jointly distributed. 

The dependence of $\Lambda^{c}$ on $c$ in (\ref{eq:most general})
encompasses the possibility of violations of free choice. If the values
of $\Lambda^{c}$ and $c$ are somehow interdependent, then different
choices of $c$ correspond to different distributions of the hidden
variable. (In Conclusion, we will discuss why it is better not to
treat $c$ as a random variable.) Note that even if the the values
of $\alpha$ do change with $c$ (i.e., $c$ is not a dummy argument),
it is still possible that the distribution of $R_{q}^{c}$ does not
depend on $c$. In other words, (\ref{eq:most general}) is compatible
with the condition of non-disturbance (non-signaling), and this is
also true for the two special cases considered next.

We say that the HVM satisfies the assumption of \emph{context-independent
(CI) mapping }(o\emph{r local causality}, when spacelike separation
is involved) if, for every $c$,
\begin{equation}
\left\{ R_{q}^{c}\right\} _{q\Yleft c}\overset{d}{=}\left\{ \beta\left(q,\Lambda^{c}\right)\right\} _{q\Yleft c}.\label{eq:locality}
\end{equation}
For system (\ref{eq:1}), this means
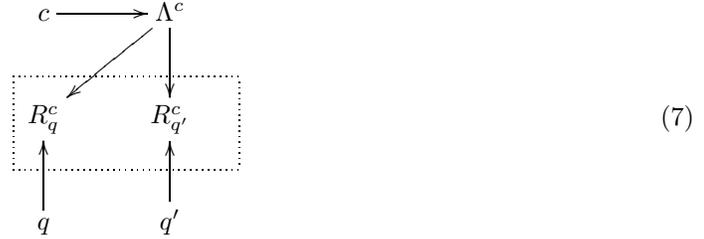
\begin{equation}
\vcenter{\xymatrix@C=1cm{c\ar[r] & \Lambda^{c}\ar[dl]\ar[d]\\
\save"2,1"."2,2"*+++[F.]\frm{}\restore R_{q}^{c} & R_{q'}^{c}\\
q\ar[u] & q'\ar[u]
}
}
\end{equation}
In the model (\ref{eq:locality}) we eliminate $c$ as an explicit
argument of function $\beta$, but generally allow $\Lambda^{c}$
to have different distributions in different contexts $c$. Clearly,
this implies that $R_{q}^{c}$ may have different distributions for
different $c$, at a fixed $q$. However, the usual view (dating back
to \citep{BellvsSHC1985}) is that CI-mapping (or local causality)
is not violated here because the dependence of $R_{q}^{c}$ on $c$
is ``indirect'': the value of $R_{q}^{c}$ at a fixed value of $\Lambda^{c}$
depends on $q$ alone. We will see below, however, that the difference
between ``direct'' and ``indirect'' dependence on $c$ is specious:
they are completely interchangeable. 

We say that the HVM satisfies the \emph{free choice} assumption if,
for every $c$,
\begin{equation}
\left\{ R_{q}^{c}\right\} _{q\Yleft c}\overset{d}{=}\left\{ \gamma\left(q,c,\Lambda\right)\right\} _{q\Yleft c}.\label{eq:free choice}
\end{equation}
In this model $\Lambda$ is one and the same for all $q$ in all contexts
$c$. In this general form, the HVM with free choice is allowed to
violate the local causality assumption. For system (\ref{eq:1}),
the model (\ref{eq:free choice}) can be presented as
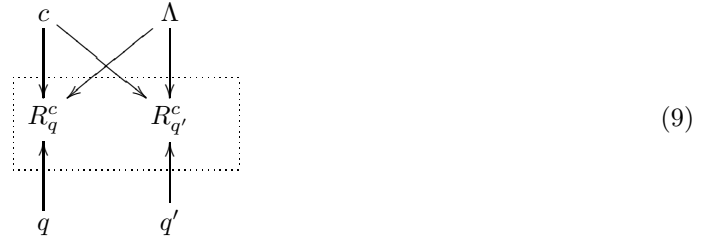
\begin{equation}
\vcenter{\xymatrix@C=1cm{c\ar[d]\ar[dr] & \Lambda\ar[d]\ar[dl]\\
\save"2,1"."2,2"*+++[F.]\frm{}\restore R_{q}^{c} & R_{q'}^{c}\\
q\ar[u] & q'\ar[u]
}
}
\end{equation}

If both the assumptions of CI-mapping and free choice are satisfied,
the HVM has the form
\begin{equation}
\left\{ R_{q}^{c}\right\} _{q\Yleft c}\overset{d}{=}\left\{ \delta\left(q,\Lambda\right)\right\} _{q\Yleft c}.\label{eq:noncontextuality}
\end{equation}
This is the HVM of a noncontextual (or locally causal) system of random
variables, one for which one derives the traditional Bell-type criteria
of noncontextuality/locality. The graph representation of this model
specialized to system (\ref{eq:1}) is
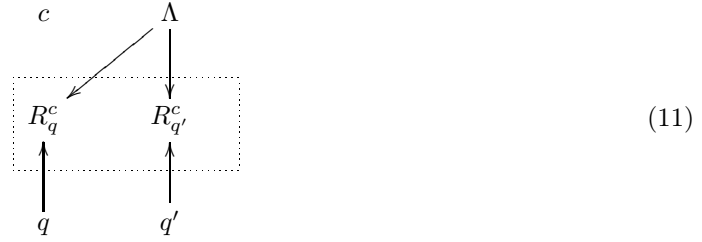
\begin{equation}
\vcenter{\xymatrix@C=1cm{c & \Lambda\ar[d]\ar[dl]\\
\save"2,1"."2,2"*+++[F.]\frm{}\restore R_{q}^{c} & R_{q'}^{c}\\
q\ar[u] & q'\ar[u]
}
}
\end{equation}

Although the only example we have given relates to the EPR/Bohm paradigm,
the definitions just given and the results below apply to all situations
described by systems of random variables, such as the classical Kochen-Specker
scenario \citep{KS1967}, the Klyachko-Can-Binicio\u{g}lu-Shumovsky
paradigm \citep{KCBS2008}, the Leggett-Garg experiments \citep{LeggGarg1985,KoflerBrukner2013},
etc., with or without the assumption of no-disturbance (or no-signaling)
\citep{DKC2020,KujDzhLar2015}.

\section{Theorem}

The main point we make in this paper is very simple and has a very
simple demonstration: HVMs (\ref{eq:most general}), (\ref{eq:locality}),
and (\ref{eq:free choice}) are pairwise equivalent. We prove this
by first showing the equivalence (\ref{eq:locality})$\Leftrightarrow$(\ref{eq:free choice}),
and then the equivalence (\ref{eq:most general})$\Leftrightarrow$(\ref{eq:free choice}). 

The proof requires one standard probabilistic notion, and one clarifying
observation. The notion in question is (probabilistic) \emph{coupling}:
given an indexed set of random variables $\mathcal{X}=\left\{ X_{i}:i\in I\right\} $,
any set of \emph{jointly distributed} random variables $Y=\left\{ Y_{i}:i\in I\right\} $
such that $X_{i}\overset{d}{=}Y_{i}$ for all $i\in I$ is called
a coupling of $\mathcal{X}$. 

The observation in question is very simple, but is sometimes misunderstood:
any indexed set of jointly distributed random variables is a random
variable in its own right, and its components can always be presented
as measurable functions of one and the same random variable. For instance,
in a vector $U=\left\{ U_{1},\ldots,U_{n}\right\} $ of jointly distributed
$\pm1$-valued variables, each $U_{i}$ can be presented as a function
of $U$ (namely, its $i$th projection $\mathrm{Proj}_{i}\left(U\right)$).
Equivalently, one can form a random variable $V$ with $2^{n}$ values,
in a bijective correspondence with $\left\{ -1,1\right\} ^{n}$, and
present each $U_{i}$ as some function $f_{i}\left(V\right)$. Another
example: if $U_{1},\ldots,U_{n}$ are jointly distributed continuous
random variables, $V$ such that $U_{i}=f_{i}\left(V\right)$ can
always be chosen to be uniformly distributed between 0 and 1 \citep{Kechris1995}.
The difference between an indexed set of jointly distributed random
variables and a ``single'' random variable is a matter of choosing
between two interchangeable representations.
\begin{proof}
\textbf{(1a)} To show that, for any $c$,
\begin{equation}
\left\{ R_{q}^{c}\right\} _{q\Yleft c}\overset{d}{=}\left\{ \beta\left(q,\Lambda^{c}\right)\right\} _{q\Yleft c}\Longrightarrow\left\{ R_{q}^{c}\right\} _{q\Yleft c}\overset{d}{=}\left\{ \gamma\left(q,c,\Lambda\right)\right\} _{q\Yleft c},
\end{equation}
we form an arbitrary coupling $\Lambda$ of the random variables $\left\{ \Lambda^{c}:c\in C\right\} $.
We have
\begin{equation}
\Lambda^{c}\overset{d}{=}\mathrm{Proj}_{c}\left(\Lambda\right)=\phi\left(c,\Lambda\right).
\end{equation}
But then
\begin{equation}
\left\{ \beta\left(q,\Lambda^{c}\right)\right\} _{q\Yleft c}\overset{d}{=}\left\{ \beta\left(q,\phi\left(c,\Lambda\right)\right)\right\} _{q\Yleft c}=\left\{ \gamma\left(q,c,\Lambda\right)\right\} _{q\Yleft c}.
\end{equation}
\textbf{(1b)} To show the reverse implication 
\begin{equation}
\left\{ R_{q}^{c}\right\} _{q\Yleft c}\overset{d}{=}\left\{ \gamma\left(q,c,\Lambda\right)\right\} _{q\Yleft c}\Longrightarrow\left\{ R_{q}^{c}\right\} _{q\Yleft c}\overset{d}{=}\left\{ \beta\left(q,\Lambda^{c}\right)\right\} _{q\Yleft c},
\end{equation}
we define, for every $c$, the random variable
\begin{equation}
\Lambda^{c}:=\left\{ \gamma\left(q,c,\Lambda\right)\right\} _{q\Yleft c}
\end{equation}
(whose components are jointly distributed because they are functions
of one and the same $\Lambda$). The components $\gamma\left(q,c,\Lambda\right)$
in $\Lambda^{c}$ are indexed by $q$, and
\begin{equation}
\gamma\left(q,c,\Lambda\right)=\mathrm{Proj}_{q}\left(\Lambda^{c}\right)=\beta\left(q,\Lambda^{c}\right).
\end{equation}
But then
\begin{equation}
\left\{ \gamma\left(q,c,\Lambda\right)\right\} _{q\Yleft c}\overset{d}{=}\left\{ \beta\left(q,\Lambda^{c}\right)\right\} _{q\Yleft c}.
\end{equation}
\textbf{(2a)} To show that, for any $c$, 
\begin{equation}
\left\{ R_{q}^{c}\right\} _{q\Yleft c}\overset{d}{=}\left\{ \alpha\left(q,c,\Lambda^{c}\right)\right\} _{q\Yleft c}\Longrightarrow\left\{ R_{q}^{c}\right\} _{q\Yleft c}\overset{d}{=}\left\{ \gamma\left(q,c,\Lambda\right)\right\} _{q\Yleft c},
\end{equation}
we again form an arbitrary coupling $\Lambda$ of the random variables
$\left\{ \Lambda^{c}:c\in C\right\} $. We have
\begin{equation}
\Lambda^{c}\overset{d}{=}\mathrm{Proj}_{c}\left(\Lambda\right)=\phi\left(c,\Lambda\right).
\end{equation}
But then
\begin{equation}
\left\{ \alpha\left(q,c,\Lambda^{c}\right)\right\} _{q\Yleft c}\overset{d}{=}\left\{ \alpha\left(q,c,\phi\left(c,\Lambda\right)\right)\right\} _{q\Yleft c}=\left\{ \gamma\left(q,c,\Lambda\right)\right\} _{q\Yleft c}.
\end{equation}
\textbf{(2b)} Finally 
\begin{equation}
\left\{ R_{q}^{c}\right\} _{q\Yleft c}\overset{d}{=}\left\{ \gamma\left(q,c,\Lambda\right)\right\} _{q\Yleft c}\Longrightarrow\left\{ R_{q}^{c}\right\} _{q\Yleft c}\overset{d}{=}\left\{ \alpha\left(q,c,\Lambda^{c}\right)\right\} _{q\Yleft c}
\end{equation}
 holds trivially. This completes the proof. 
\end{proof}
The parts (2a) and (2b) imply that the dependence of the hidden variable
on context $c$ in $\alpha\left(q,c,\Lambda^{c}\right)$ is superfluous:
it can always be eliminated by replacing $\alpha\left(q,c,\Lambda^{c}\right)$
with $\gamma\left(q,c,\Lambda\right)$. The parts (1a) and (1b) show
that the ``indirect'' dependence of $\beta\left(q,\Lambda^{c}\right)$
on $c$ is not a special form of dependence: any $\gamma\left(q,c,\Lambda\right)$
or $\alpha\left(q,c,\Lambda^{c}\right)$ can be presented as $\beta\left(q,\Lambda^{c}\right)$. 

An unexpected if not paradoxical consequence of the theorem is that
an HVM of the form (\ref{eq:locality}), in spite of being introduced
as one satisfying CI mapping, does not in fact in fact impose any
constraints on the HVMs that can describe the same system of random
variables. In particular, (\ref{eq:locality}) does not prevent this
system of random variables from being modeled by an HVM of the form
(\ref{eq:free choice}), one in which CI mapping generally does not
hold. Conversely, an HVM of the form (\ref{eq:free choice}), in spite
of being introduced as one satisfying the free choice assumption,
does not in fact prevent the same system of random variables from
being modeled by an HVM of the form (\ref{eq:locality}), one in which
free choice generally does not hold. This observation provides a simple
explanation for the long since noticed reciprocity between measures
of the degree to which an HVM violates the assumptions of CI mapping
and free choice (as discussed in Conclusion).

Another implication of the same observation is that if one rejects
the possibility that a system of random variables can be described
by an HVM containing context $c$ as one of its non-dummy arguments,
then the possible HVMs for this scenario are of the form (\ref{eq:noncontextuality})
rather than (\ref{eq:locality}). This seems to have been John Bell's
original idea, criticized in \citep{BellvsSHC1985}.

\section{Discussion}

Here we discuss a few questions that can be raised in response to
the foregoing theorem.

(1) Would the analysis change if $\alpha\left(q,c,\Lambda^{c}\right)$
in (\ref{eq:most general}) were replaced with a seemingly more general
$\xi\left(q,c,\Lambda_{q}^{c}\right)$? The answer is it would make
no difference, because by forming the random variable $\Lambda^{c}:=\left\{ \Lambda_{q}^{c}\right\} _{q\Yleft c}$,
we get
\begin{equation}
\xi\left(q,c,\Lambda_{q}^{c}\right)=\xi\left(q,c,\mathrm{Proj}_{q}\left(\Lambda^{c}\right)\right)=\alpha\left(q,c,\Lambda^{c}\right).
\end{equation}
The reason the components of $\left\{ \Lambda_{q}^{c}\right\} _{q\Yleft c}$
are jointly distributed is that so are $R_{q}^{c}$ for any given
$c$.

(2) Could not the free choice assumption be violated in the form 
\begin{equation}
\left\{ R_{q}^{c}\right\} _{q\Yleft c}\overset{d}{=}\left\{ \varrho\left(q,\Lambda^{q}\right)\right\} _{q\Yleft c},
\end{equation}
clearly compatible with CI-mapping (or local causality)? The answer
is it would make no difference, because $\varrho\left(q,\Lambda^{q}\right)$
can always be written as $\delta\left(q,\Lambda\right)$. Indeed,
considering $\Lambda$ as the coupling of all $\Lambda^{q}$ such
that $q\Yleft c$ for some $c$, we get
\begin{equation}
\varrho\left(q,\Lambda^{q}\right)\overset{d}{=}\varrho\left(q,\mathrm{Proj}_{q}\left(\Lambda\right)\right)=\delta\left(q,\Lambda\right).
\end{equation}

(3) Does not the difference between $\beta\left(q,\Lambda^{c}\right)$
and $\gamma\left(q,c,\Lambda\right)$ lie in the physical meaning
of the dependence of these functions on $c$, in spite of their mathematical
equivalence? The answer is negative once again. Not having any way
to observe $\Lambda$, we can impose no physical constraints on what
it is and how it can cause changes in the measurement outcomes $R_{q}^{c}$.
Thus, for contexts with spacelike separated components, whatever $\Lambda$
is, if one adopts the local causality assumption, $\Lambda$ cannot
transfer information from spacelike remote components of $c$ to $R_{q}^{c}$.
If the dependence of $\Lambda$ on the remote components of $c$ is
explained by a common cause in their inverted light cones, then the
same explanation applies to $R_{q}^{c}$ and $c$ directly, leading
to the representation (\ref{eq:free choice}). 

(4) Should not the free choice assumption be formulated in terms of
the (non-)independence of the hidden variable $\Lambda$ and context
$c$ treated as another random variable? The answer is that, first,
this makes no difference, and second, treating $c$ as a random variable
is conceptually dubious. The reason this makes no difference is that,
even if $c$ is a random variable such that $c$ and $\Lambda$ are
jointly distributed, conditioning $\Lambda$ on different values of
$c$ creates the variables $\Lambda^{c}$ of the analysis above. The
reason why treating $c$ as a random variable is dubious is that one
can easily realize experimental procedures in which $c$ is not chosen
randomly: e.g., one can run four side-by-side EPR/Bohm experiments,
each with a fixed value of $c$ for years; or one can change the values
of $c$ in accordance with a deterministic algorithm (say, $1,2,3,4,1,2,3,4,\ldots$).
The notions of randomness and of a random variable are not identical,
so the procedures just mentioned may still allow one to treat $c$
as a random variable --- but this cannot be done uniquely, in a standard
way based on frequencies of occurrences. Conditioning $\Lambda$ on
$c,$ on the other hand, is innocuous: realizations of $\Lambda$
cannot be controlled in any way.

\section{Conclusion}

The analysis presented in this article has consequences beyond just
the issue of how one derives the Bell-type criteria. There is a sizable
body of insightful literature on the reciprocity between the degree
free choice is violated and the degree of dependence of measurements
on contexts \citep{BarrettGisin2011,Hall2011,Rossetetal.2014,Friedmanetal2019}.
It is clear now that the logical basis of this literature, usually
focusing on specific cases, such as the EPR/Bohm type experiment,
is the equivalence of (\ref{eq:free choice}) and (\ref{eq:locality}).
We know now that this equivalence holds in complete generality, for
all possible systems of random variables (\ref{eq:system}), with
and without disturbance alike. This equivalence implies, in particular,
that any measure of deviation of an HVM from the model (\ref{eq:noncontextuality})
should equally be interpretable as the degree of violation of CI mapping
(local causality) and the degree to which the experimenters lack free
choice. This is most clearly indicated in the recent paper by Blasiak
et al. \citep{Blasiaketal.2021}. The abstract of their paper states
that ``causal explanations resorting to either locality or free choice
violations are fully interchangeable.'' This coincides with the results
of the present work, except that here they are established by a very
different argument and in greater if not maximal generality. 

\paragraph{Acknowledgments}

I am grateful to Matthew Jones for critically discussing with me earlier
versions of this paper. 

\paragraph{Conflict of Interests}

Author declares no conflicts of interests

\paragraph{Data Availability}

Data sharing not applicable to this article as no datasets were generated
or analyzed during the current study.

\end{document}